\documentclass{aa}

\usepackage[dvips]{graphicx}
\usepackage{natbib}
\usepackage[latin1]{inputenc}

\usepackage[usenames]{color}

\begin{document}

\title{Relation between trees of fragmenting granules  and supergranulation evolution}

\author{ Th.~Roudier\inst{1}, J.M.~Malherbe\inst{2}, M.~Rieutord\inst{1}, Z. ~Frank\inst{3}}

\date{Received \today  / Submitted }

\offprints{Th. Roudier}

\institute{Institut de Recherche en Astrophysique et Plan\'etologie, Universit\'e de Toulouse, CNRS,
14 avenue Edouard Belin, 31400 Toulouse, France
\and LESIA, Observatoire de Paris, Section de Meudon, 92195 Meudon, France
\and Lockheed Martin Advance Technology Center, Palo Alto, CA, USA}

\authorrunning{Roudier et al.}
\titlerunning{Relation between tree of fragmenting granules (TFG) and supergranulation evolution from Hinode observation}

\abstract{
  The determination of the underlying mechanisms of the magnetic elements diffusion over the solar surface is still a challenge.
 Understanding the formation and evolution of the solar network (NE) is a challenge, because it provides a magnetic flux over the solar
 surface comparable to the flux of active regions at solar maximum.
}{
  We investigate the structure and evolution of interior cells of solar supergranulation. From Hinode observations, we
 explore the motions on solar surface at high spatial and temporal resolution. We derive the main organization
 of the flows inside supergranules and their effect on the magnetic elements.
}{
  To probe the superganule interior cell, we used the Trees of Fragmenting Granules (TFG) evolution and their relations 
to horizontal flows.
}{
 Evolution of TFG and their mutual interactions result in cumulative effects able to build horizontal coherent
flows with longer lifetime than granulation (1 to 2 hours) over a
scale up to 12\arcsec. These flows clearly act on the diffusion of the intranetwork (IN) magnetic elements 
and also on the location and shape of the network.
}{
From our analysis during 24 hours, TFG appear as one of the major
elements of the supergranules which diffuse and advect the magnetic
field on the Sun's surface. The strongest supergranules contribute the most 
to magnetic flux diffusion in the solar photosphere.
 }

\maketitle

\section{Introduction}

The magnetic field structure of the quiet solar surface is still
puzzling. Very recent observations changed the paradigm of the magnetic 
flux evolution in the quiet network. For instance, \citet{GBOKD14} have shown 
that, from very high quality magnetograms obtained with the Narrowband Filter 
Imager (NFI) of the Solar Optical Telescope (SOT) onboard the Hinode satellite, 
most of the magnetic flux of the network comes from the weak part that is produced 
inside the supergranulation cells. More precisely, they observed that $38\%$ of 
the flux that appears inside supergranular cells moves toward the frontiers and 
interacts with the magnetic field of the network. 
The quiet-Sun magnetic elements appear to have two distinct velocity components, 
one random and one systematic, both depending on the location of the element 
within the supergranular cell \citet{OBK2012}. The difficulty of describing the 
dynamics of the magnetic flux in the quiet-Sun is to a great part due 
to the difficulty of characterizing the supergranule properties and evolution 
and of following small-scale magnetic elements. The supergranules have been studied by many
authors, for instance, by \citet{H54}, \citet{LNS62}, see also the review by
\citet{RR10}. As underlined in this last paper, the reported properties
of supergranulation depend on the procedure used in the data
processing. Many supergranular characteristics remain unclear. We still lack an observational 
description of the appearance of supergranules and their evolution (see \cite{STW95}).
 The question that are still open are for instance the difficulty in measuring the temperature 
difference between the center and the edge of supergranules this should be straightforward 
if they are convective cells. Another question is why the vertical component of supergranules
been found at mesoscales \citet{Nov1989}. At the supergranular
scale, the magnetic network is quite fragmented, corresponding to
patches of strong magnetic fields associated with vertices where flows converge 
\citet{CCT07}.  The network is rarely closed at supergranule edges and the mechanism that cause this
is still not understood.  Answers to these questions
are fundamental for understanding the physics of superganules and their 
interactions with magnetic elements.  The vertical magnetic fields, which are commonly 
considered as passive scalars, are transported toward supergranular boundaries, 
but the feedback of the field on the flow is unknown. Many questions
remain about the origin of the supergranulation: superganulation might be a surface manifestation 
of interior motions  or supergranules might result from interaction of granules, 
or be related to deep convection or even a consequence of magneto-convection \citet{HS2014}. 
The scale (20 to 30 Mm) and the long lifetime (24 to 48 hours) of the supergranulation has caused small 
telescopes or large-scale (spatial and temporal) smoothing windows to be used but these methods do 
not permit describing the fine structure of supergranules and that of the physical processes that govern
the supergranular scale.

In the present paper, trees of fragmenting granules (TFG) are used to probe the interior of 
supergranules and their relationship with the magnetic field evolution.
A TFG consists of a family of repeatedly splitting granules, that originate from a single granule 
in the beginning. In Sect. 2 we describe the observations and data reduction. Section 3 considers 
with horizontal velocities inside supergranules. 
The link between exploding granules and flow divergence is discussed in Sect. 4.
The relations between TFG and horizontal velocities are studied
in Sect. 5. In Sect. 6 we investigate the evolution of the intranetwork and corks. 
The influence of the strongest horizontal flows on the diffusion and advection of the
magnetic elements is presented in Sect. 7. We discuss the role of the TFG on the
diffusion of magnetic elements in Sect. 8. In conclusion, we emphasize
the role of TFG in the diffusion of intranetwork magnetic elements
and the formation of the network.

 \section{Observation and data reduction}

We used multiwavelength data sets of the Solar Optical Telescope
(SOT)  onboard the Hinode mission (Ichimoto et al., 2004;
\citealt{STISO08}). The SOT has a 50 cm primary mirror with a
spatial resolution of about 0.2\arcsec at 550 nm. For our study, we
used multiwavelength observations from the SOT  NarrowBand
Filter Imager (NFI) and Broadband Filter Imager (BFI). The SOT 
measures the Stokes parameters I and V of the FeI line at 630.2 nm
with a spatial resolution 0.16 \arcsec. More precisely, the Lyot
filter is set to a single wavelength in the blue wing of the line,
typically 120 mÅ for Fe I magnetograms. Then images are taken at
various phases of the rotating polarization modulator and are added to 
 or subtracted from a smart memory area in the onboard computer. The BFI
scans consist of time sequences obtained in the blue continuum
(450.4 nm), G band (430.5 nm) and Ca II H (396.8 nm). The
observations were recorded continuously from 29 August 10:17 UT to
31 August 10:19 UT 2007, except for an interruption of 7 minutes at disk
center on 30 August at 10:43 UT. The solar rotation is compensated
for to follow exactly the same region on the Sun. The time step
between two successive frames is 50.1 s. The field of view (FOV) of BFI
images is $111. 6~\arcsec \times 111. 6~\arcsec$ with a pixel size
of 0. 109~\arcsec ($1024\times1024$). After alignment, useable FOV is 
reduced to $100~\arcsec \times92~\arcsec$. To
remove the effects of oscillations, we applied a subsonic Fourier
filter. This filter is defined by a cone in the $k-\omega$ space,
where k and $\omega$ are spatial and temporal frequencies. All
Fourier components of $\omega/k$  $ \leq  6 ~\rm km~s^{-1}$ were
retained to keep only convective motions (Title et al. 1989). To
detect TFG, granules were labeled in time as described in
\citet{RLRBM03}.

We extracted a subfield (60~\arcsec x 62~\arcsec) of our data
centered on the well-formed supergranule bounded by an almost closed
magnetic network, centered on (24~\arcsec, 39~\arcsec) in
Fig.~\ref{vit}. This subfield is called the supergranule field.

 We built several movies, provided as electronic supplemental material (ESM), using the
1716 time steps that show the evolution of granule families, horizontal
velocities, divergences, longitudinal magnetic field and other
related quantities (see the appendix for more details).

\section{Horizontal motions derived from local correlation tracking}

 Horizontal velocities were derived from the  local correlation tracking (LCT) (\citealt{RRMV99}) using
a temporal window of 30 minutes and a spatial window of 3 arcsec
FWHM. Hence we have 48 fields of $v_{x}, v_{y}$ values along the 24-hour sequence of the 
supergranule field. From the horizontal velocity vector 
($v_{x}, v_{y}$), measured by the LCT, the magnitude ($Vh\_mag=\sqrt{{v_{x}^{2}}+{v_{y}^{2}}}$) 
was derived. The median value of velocities plotted in Fig.~\ref{vit} is 0.4 km $s^{-1}$, 
twice lower than that of magnetic elements that we measured directly (see below). 
This result is expected, because velocities are smoothed by the windows chosen for the LCT (in
particular the spatial window is much larger than the size of
magnetic elements). When we compared between LCT velocities and true plasma 
velocities, performed using results of a Stein magnetohydrodynamic (MHD) simulation (\citealt{Stein2009}), 
we also found a difference by a factor of two. Figure~\ref{ang} shows the
angular distribution of the velocity vector with respect to the
radial direction from the supergranule center (24~\arcsec,
39~\arcsec), which shows the main role of outward motions. The half-height of the Gaussian fitting 
corresponds to departures of $\pm45°$ from the radial direction.

\begin{figure}
\centering
\includegraphics[width=9cm]{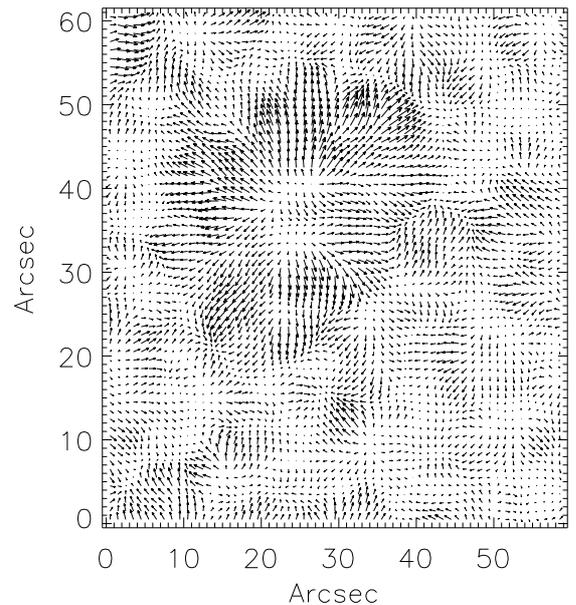}
 \caption[]{Horizontal velocity at time 12 hours, spatial and temporal window 3 \arcsec
and 30 min. The supergranule field is centered on~(24~\arcsec, 39~\arcsec)} \label{vit}
\end{figure}

\begin{figure}
\centering
\includegraphics[width=9cm]{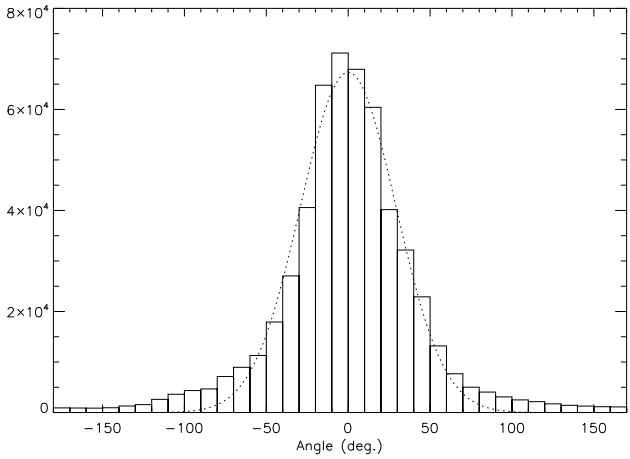}
 \caption[]{Histogram of horizontal velocity directions deduced from the LCT
 (solid line). A Gaussian fit (dotted line) is superimposed.}
\label{ang}
\end{figure}

\section{Link between exploding granules and large amplitude divergence.}

An important point to elucidate is the origin of large-scale
positive divergences in the TFG and their link with horizontal
velocity flow.

\begin{figure}
\includegraphics[width=9cm]{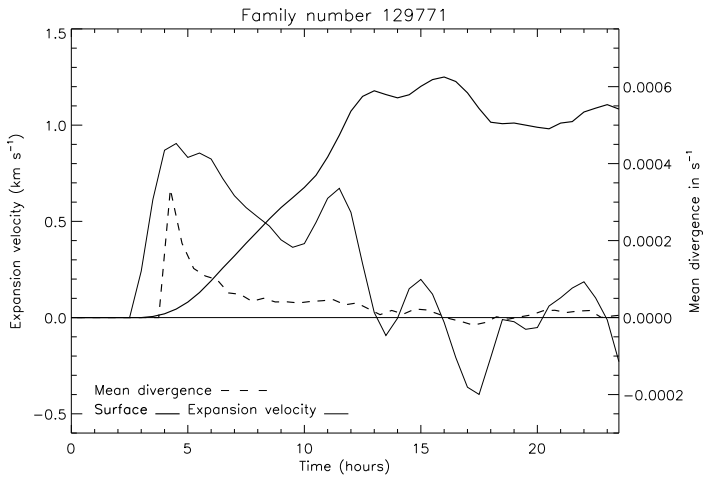}
\caption[]{Area of the largest TFG (number 129771, solid line), mean velocity divergence 
(dashed line) and expansion velocity (dotted line).
} \label{divarea}
\end{figure}

\begin{figure}
\includegraphics[width=9cm]{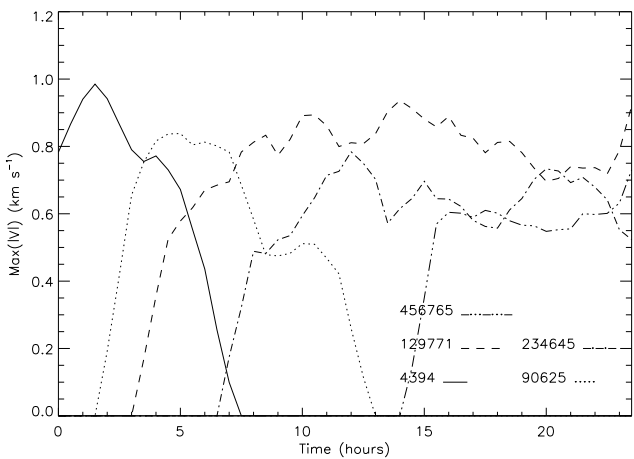}
\caption[]{ Highest horizontal velocity of the five largest TFGs in
the supergranule. } \label{Horvit}
\end{figure}

\begin{figure}
\includegraphics[width=9cm]{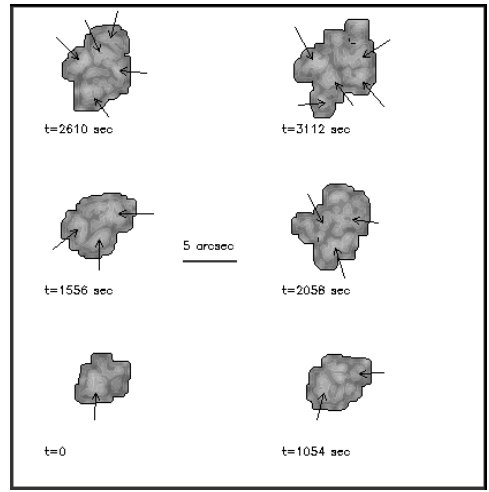}
\caption[]{Evolution of a new family of granules (TFG number 129771)
where exploding granules are indicated by arrows. The dark bar gives the scale of 5~\arcsec } \label{explogra}
\end{figure}

\begin{figure}
\includegraphics[width=9cm]{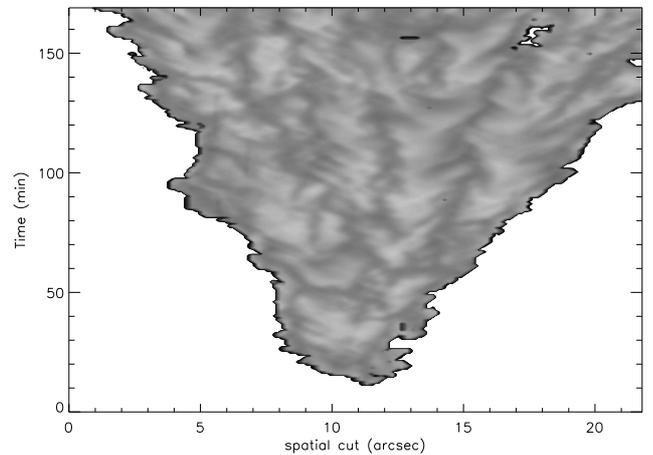}
\caption[]{V-shape temporal slice of successive exploding granules
of a TFG (number 129771) } \label{explocut}
\end{figure}

Figure~\ref{divarea} shows family of the supergranule (number 129771) that is the strongest in space and time.
It develops from time t = 3 hours of the sequence and covers the greatest extent of the solar surface at 
t = 13 hours after which it remains approximately stationary. 
The average value of the velocity divergence is always positive and highest at the beginning of the
family development. The expansion velocity (defined as the time
derivative of the square root of the family surface) is also highest 
(0.9 km/s) at the beginning, meanwhile the maximum horizontal
velocity also attains 0.9 km/s. It is important to note that high
divergences and velocities continue to be observed after the family has developed. 
This means that granules explode permanently in the family.

In Fig~\ref{Horvit}, the highest horizontal velocity is measured in
the five strongest families of the supergranule.  The largest family is number 129771  
and reaches 1.0 km/s. Families  often compete when they develope together. Maxima appear at times t =
1.5, 5, 12, 14 and 17 hours where each maximum contributes to the 
migration of magnetic elements toward the boundaries. Their maxima
are not in phase but in sequence, allowing a collective action in
time to spread the magnetic field across the solar surface toward the
network.

\subsection{Birth and evolution of TFG}

 The TFG evolution  shows successive explosions of granules at the beginning. 
At this stage, these explosions appear
quasi-simultaneous. Figure~\ref{explogra} displays a family evolution
at each step of a new generation of the tree (the time  origin is
the time of the first granule explosion of the TFG); arrows indicate
the location of the observed explosions. From one generation to the
next, the number of exploders may increase. The quasi-simultaneity
of explosions is remarkable until up to five successive generations; then
explosions shift from one generation to the next  as a result of small
cumulative phase shifts observed from the beginning.
Figure~\ref{explocut} shows a temporal slice of a TFG (the same as in Fig~\ref{explogra}) 
where the V shape indicates the location of
exploding granules. This shows the link between
numerous exploding granules and the development of the TFG in the
first three hours.

 The time difference between two successive explosions is generally in the range of 8 to 10 minutes,
then the granule children explode more or less in phase at the same
time. This quasi-simultaneity of explosions at the beginning of a
new TFG produces collective effects on the TFG expansion but also on the 
horizontal velocity fields. In particular, we observe
(Fig~\ref{divarea}) a broad divergence of the flow at the location
of the new TFG which later decreases. The largest area in
Fig~\ref{divarea} (around 13:00 and 16:00) is related to an increase
of 73\% of exploding granules in the TFG relative to the minimum
at 14:30.

During its lifetime, the TFG shows variations related to new
coherent exploding granules in the tree, which create new branches.

\subsection{Relation of divergence and horizontal velocity}

\begin{figure*}
\centering
\includegraphics[width=9cm]{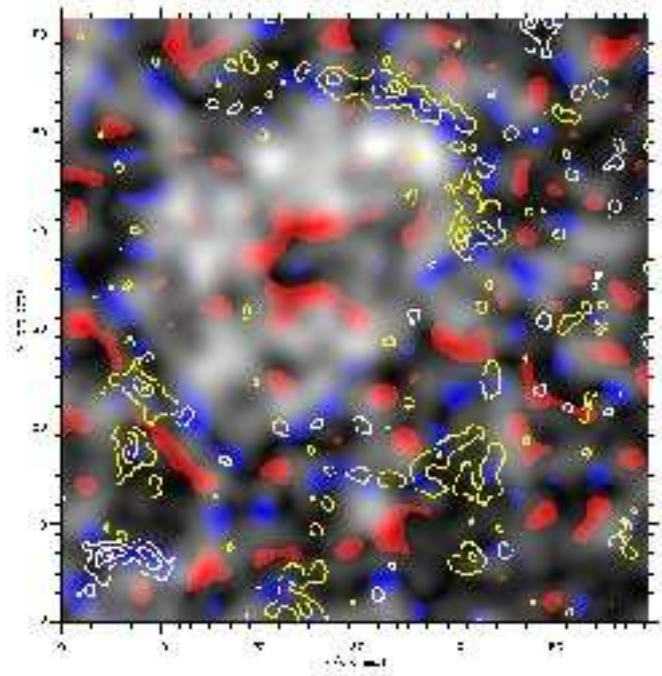}
\caption[]{{\bf Evolution of the properties of a supergranular cell. An animation is 
available. See the appendix for details.}~ Horizontal velocity amplitude (grayscale) 
in the sub-FoV. The divergence field is represented in red (positive) and blue (negative). The 
LoS magnetic field contour is shown in yellow and white (positive and negative polarities). 

}\label{V_divV_B}
\end{figure*}

  The divergence field (length scale between 2 and  7 arcsec) is shown in
Fig~\ref{V_divV_B} together with horizontal velocities $Vh\_mag$ and
longitudinal magnetic fields at time 05:36 of the sequence. The
strongest $Vh\_mag$ are located on the edge of divergences showing a
clear link between them and horizontal flows inside the
supergranule. The divergence in the central part of the FOV issued
from a new TFG (23\arcsec, 34\arcsec) exhibits high $Vh\_mag$ in its
lower part, although in its upper part (10\arcsec, 48\arcsec) only
low $Vh\_mag$ is observed. This asymmetry is due to the presence of
another divergence produced by a neighboring TFG (28\arcsec,
40\arcsec). Horizontal flows in between are opposite in direction,
resulting in weak velocity flow that is  due to mutual interaction. 
In contrast, TFG interactions may add to generate large-scale
$Vh\_mag$ (few arcsec) which are important to drag the magnetic
elements across the solar surface.

Panel 6 of Fig~\ref{V_divV_B} shows the temporal evolution of $Vh\_mag$, divergences, 
and magnetic fields during 24 hours in the same FOV as panel 1.
First we note that NE magnetic fields are located in the converging
flows that delineate the supergranule. This movie shows the links
between positive divergences and $Vh\_mag$. Strong $Vh\_mag$ occur
where no opposite flow is created by other divergences in the
vicinity. Successive divergences may have a cumulative effect on the
final $Vh\_mag$ amplitude. Then the space and time coherence of the
divergence appears crucial for the magnitude on the horizontal flow.
Collective effects reinforce the $Vh\_mag$ field. Observed $Vh\_mag$
are in phase on scales of several arcsec and propagate velocity
fronts across the solar surface.

In panel 7 the strongest divergences are located in the family; at the
TFG birth, positive divergence is observed and the highest
divergence magnitude corresponds to new branches of the TFG that 
grow quickly.

Panel 8 shows the relation between the strongest positive divergence,
$Vh\_mag$,  and the families.  It allows us to describe the evolution of
the family and divergences. At the emergence of a new family, a new
positive divergence arises and generates strong horizontal flows
which then propagate across the surface. During their travel, new
flows combine with preexisting flows and thus generate the observed
$Vh\_mag$, which is modulated by collective effects.

\subsection{Model of the collective effect of exploding granules}

\begin{figure}[h]
\centering
\includegraphics[width=5cm]{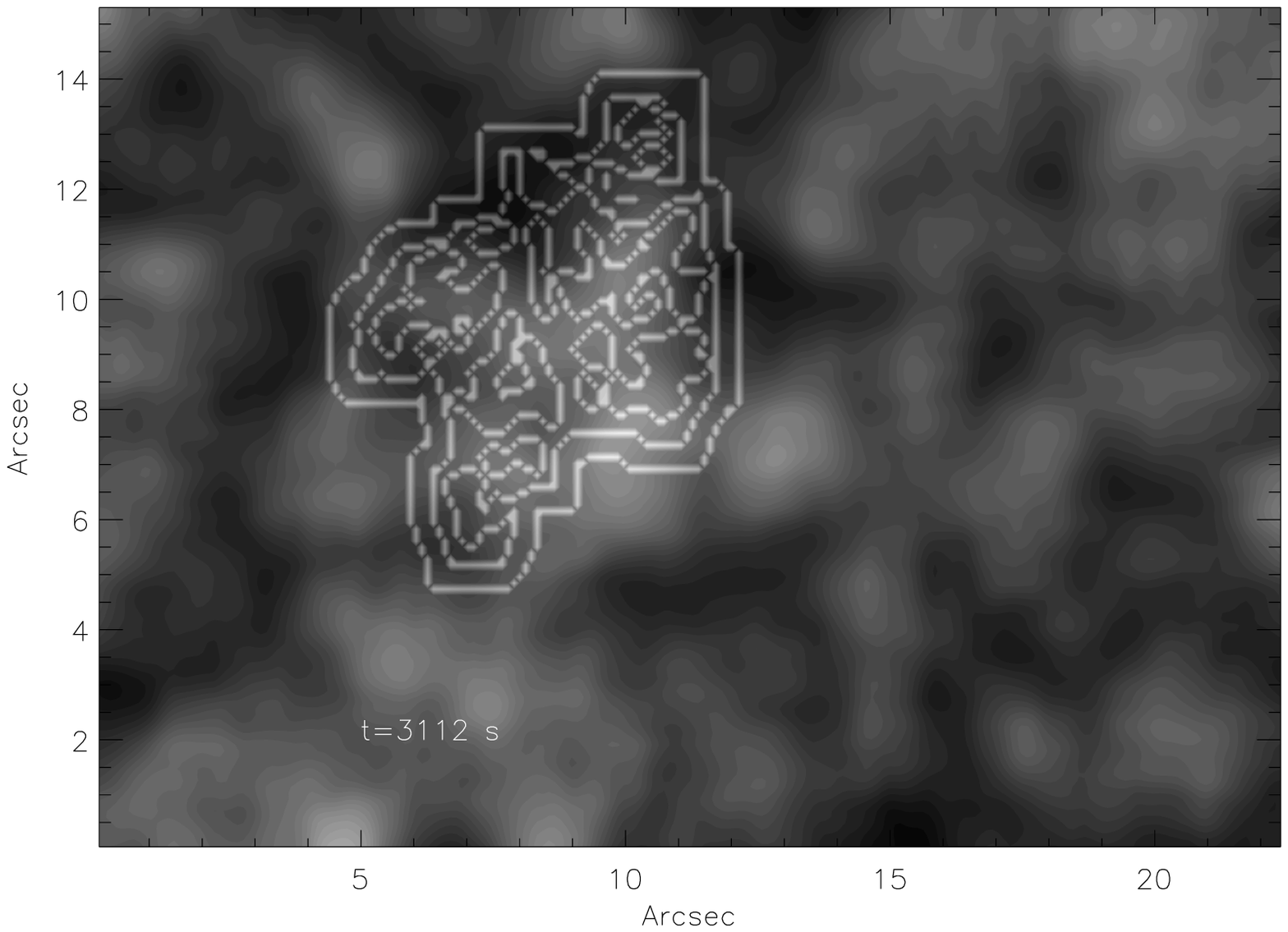}
\includegraphics[width=5cm]{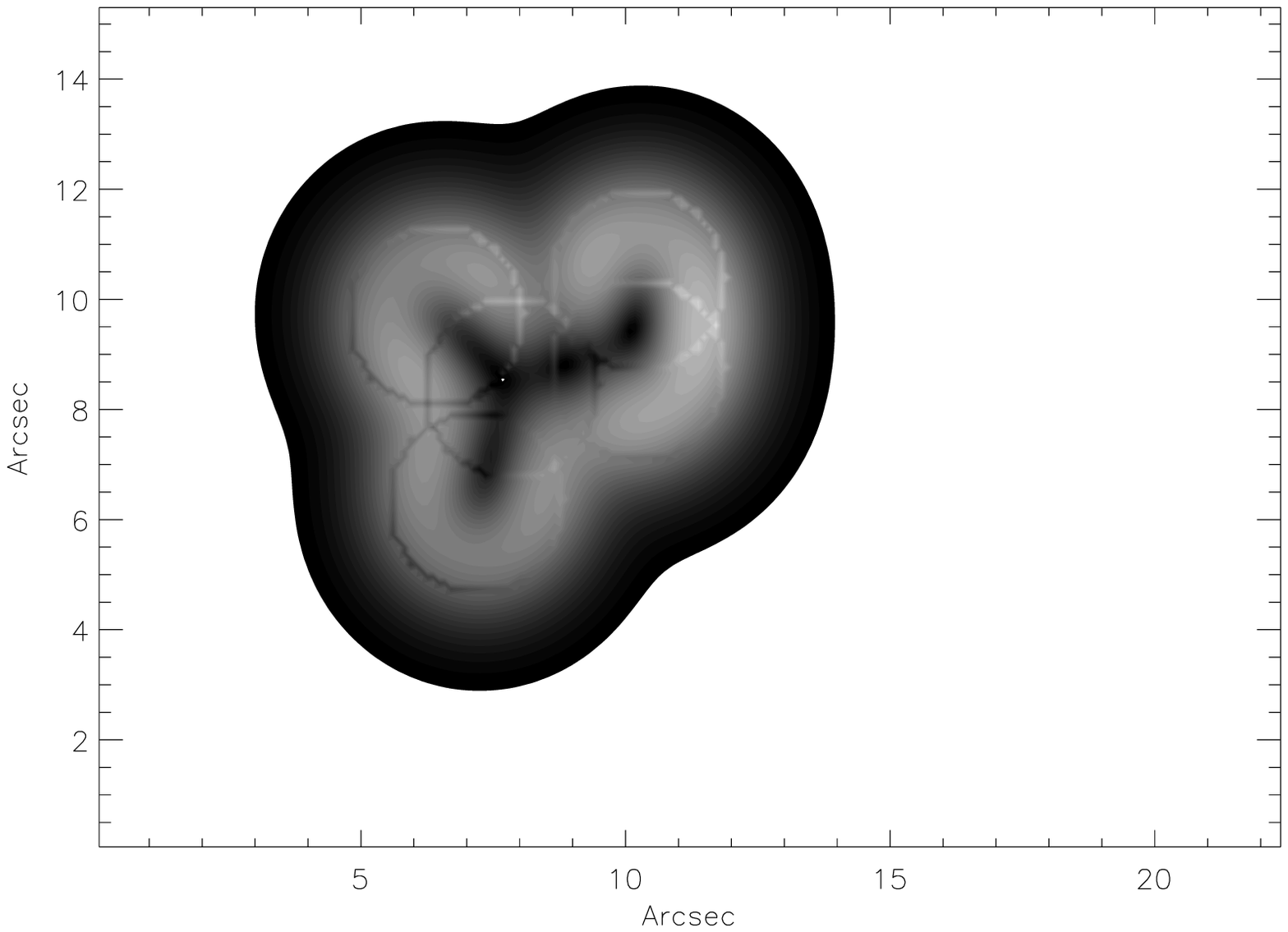}
\caption[]{Horizontal velocity amplitude field (gray levels) overlapped by intensity contour of 
the observed  five exploding granules (top). Horizontal velocity amplitude (gray levels) of the five simulated 
exploding granules (bottom)} 
\label{explofive}
\end{figure}

\begin{figure*}
\centering
\includegraphics[width=12cm]{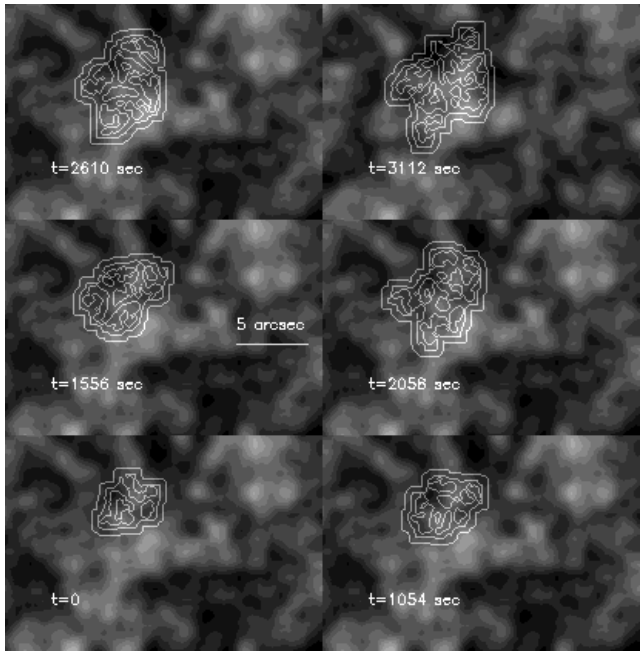}
\caption[]{Evolution of the exploding granules (contours) and
amplitude of horizontal velocities (gray levels). The white bar gives a scale of 5~\arcsec.} 
\label{explogra2}
\end{figure*}

 To model the collective effect of exploding granules on the flow, we simulated exploding granules
by the function $V(r)= Vexp\times (r/R) \times \exp(-(r/R)^2)$, where
Vexp is  the amplitude  velocity,  R the radius of the granule and r
the distance from the granule center. Figure~\ref{explofive} shows the
comparison of horizontal velocity magnitudes issued from the 
simultaneous explosion of five granules and compare them with those
observed at t = 3112 sec on the Sun. In this simulation the
parameters are Vexp = 1.33 km/s and R = 3000 km. The simulation
forms the highest velocity at the edge of the group of five exploding
granules.  The simultaneous explosions produce a collective effect
on the flow with length greater than the size of granules up to
several arcsec. Figure~\ref{explogra2} shows the temporal evolution of
the outlined intensity of the granules superimposed on the  horizontal
velocities $Vh\_mag$. This confirms that larger amplitudes of velocities are 
at the edge of the family where explosions are strongest.
However, departures between observed and simulated highest velocities might 
also be explained by the difference in the expansion rate of the
solar explosive granules (which was chosen to be uniform in the
simulation) or by the small phase shift between explosions of
granules, which are neglected in the model.

\section{Dynamical properties of the TFG and relation with horizontal velocity magnitude}

\begin{figure*}
\centering
\includegraphics[width=24cm]{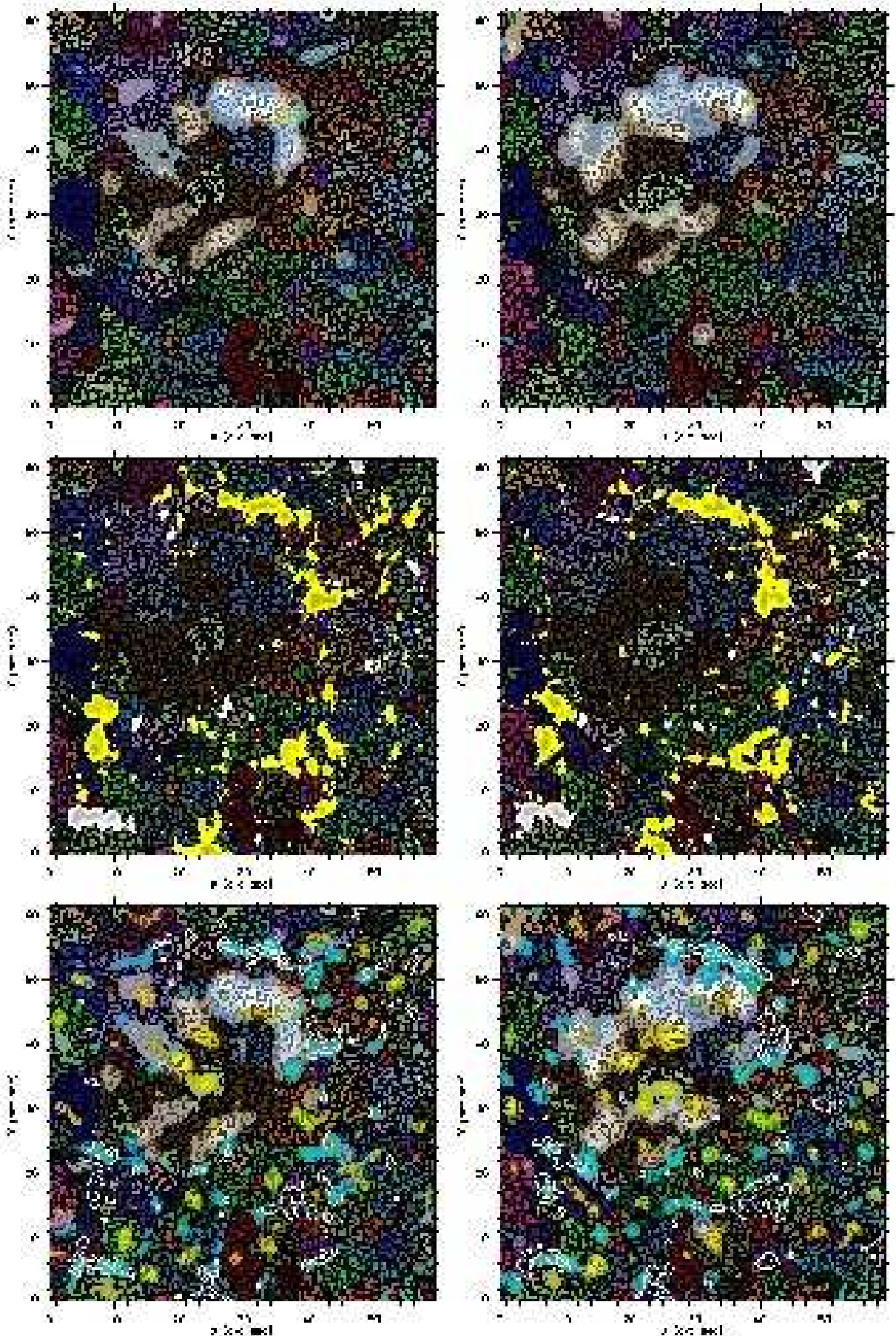}
\caption[]{{\bf Evolution of the properties of a supergranular cell. An animation is 
available online. See the appendix for details.}~ (Top left) Horizontal velocity magnitude $Vh\_mag$ (gray
levels) and families (TFG) at time 05:02; (top right) same plot at
05:36; families are represented by various individual colors.
(Middle left) Longitudinal magnetic field (yellow and white) and
families (TFG) at time 05:02; (middle right) same plot at 05:36.
(Bottom left) Horizontal velocity magnitude $Vh\_mag$ (gray levels),
families (TFG), absolute value of the longitudinal magnetic field (white
isocontours) and velocity divergence (blue for negative, yellow for
positive) at time 05:02; (bottom right) same plot at 05:36.}
\label{vit_B2}
\end{figure*}

The main goal of the present section is to determine how and where
the velocity maxima $Vh\_mag$ are produced.  Little is known about
the dynamics on the intermediate timescales between one to
several hours. Our preceding analysis \citet{RLRBM03} indicated that
solar surface is structured by families of granules (TFG) that 
originate from a granule genitor. We observed that most of the
granules on the solar surface belong to long-lived TFG. Each TFG is
characterized by properties such as lifetime, size and expansion rate.
The final area covered by the TFG seems to increase as a function of
lifetime (Fig. 7 in \citealt{RLRBM03}).The largest TFGs are close in
size to the smallest supergranules and we found that exploding
granules play a major role in TFG evolution. In particular, we
demonstrated that increases of the total area of families is related
to granule explosions. A TFG can be viewed schematically as a series 
of branches corresponding to new generations of granules.

The evolution of horizontal velocities over a large FOV during 24 hours
reveals the locations of maxima of $Vh\_mag$ that coincide with the
edges of the nascent TFG (see panel 9).

In panel 10 we observe the birth of different families; 
their expansion is associated with maxima of $Vh\_mag$ at the edge of TFGs
and the motions of these maxima look like a wave  propagating
outward. Figure~\ref{vit_B2} (top) shows an example of such an evolution.
In the central part of the figure a new TFG (green) forms around
05:02 and then expands; at 05:36, 32 minutes later, a maximum of
horizontal velocity $Vh\_mag$ appears at the bottom edge of the TFG.

Panel 10 shows that the propagation of maxima is affected
first by the growth of the family that initiated these maxima but
also by the influence of other families. In some regions, $Vh\_mag$
maxima from different TFG may combine to form larger scale
$Vh\_mag$ structures (see Panel 5).

Figure~\ref{vit_B2} (middle) shows that longitudinal magnetic fields
are located at the edge of the TFG (both NE and IN). $Vh\_mag$,
families and the magnetic fields are shown together in
Fig~\ref{vit_B2} (bottom) and panel 8; we can follow the evolution
of the large TFG (brown, in the center of the field) and $Vh\_mag$
maxima close to its borders. At 05:02, the new green TFG competes 
with the brown TFG . The maxima of $Vh\_mag$ around
these two families are clearly visible at 05:40 of the 24-hour
sequence. The expansion speed of these two TFG vanishes along their
common frontiers and causes an almost circular $Vh\_mag$ shape around
the brown TFG on a scale between 10 to 15 arcsec. This is a good
example of $Vh\_mag$ maxima generation by the evolving combination
of strong adjacent TFG.

The entire sequence shows that the evolution of $Vh\_mag$ maxima is 
intimately related to the life of TFG through their location, strength, 
and birth date. Panels suggest a constant production rate of
$Vh\_mag$ structures, with magnitudes related to  the competing TFG 
in the FOV, but also to existing $Vh\_mag$ areas that are due to previous TFG that 
decline or have just disappeared. The frequent occurrence of
$Vh\_mag$ patches and the interaction between families produces several events 
that contribute to the diffusion of the magnetic field and
the photospheric network in the quiet Sun.

\section{Intranetwork (IN) magnetic elements and cork evolution}

\subsection{Intranetwork magnetic elements}

\begin{figure}
\centering
\includegraphics[width=9cm]{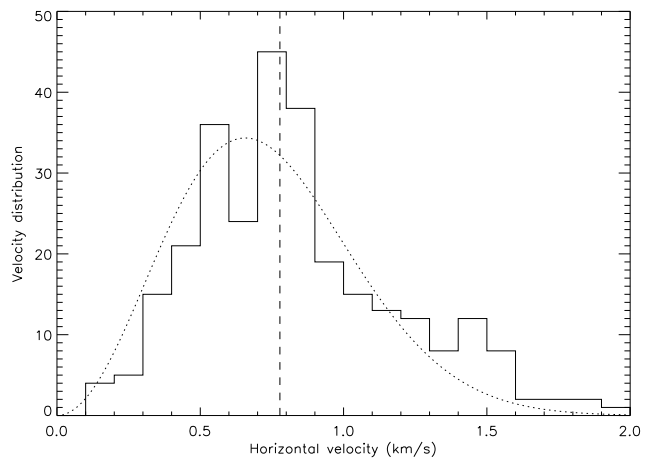}
 \caption[]{Histogram of horizontal velocities for observed magnetic elements 
inside the supergranule (solid line). A Maxwellian fit (dotted line) is superimposed.}
\label{elvel}
\end{figure}

\begin{figure}
\centering
\includegraphics[width=9cm]{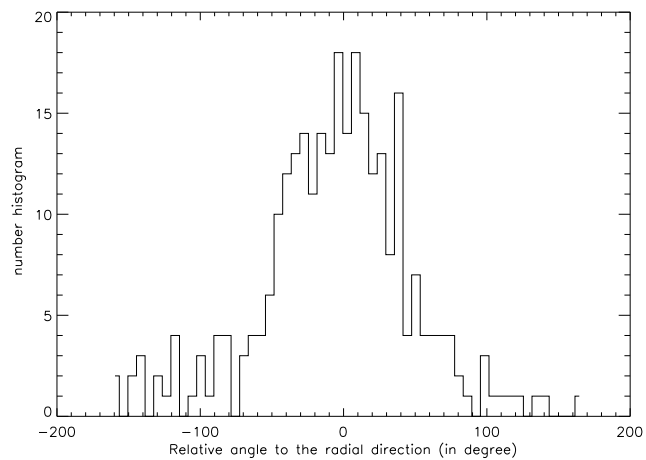}
 \caption[]{Histogram of the angular distribution of the velocity vector of
the IN magnetic elements with respect to the radial direction.}
\label{INelement}
\end{figure}

\begin{figure}
\centering
\includegraphics[width=9cm]{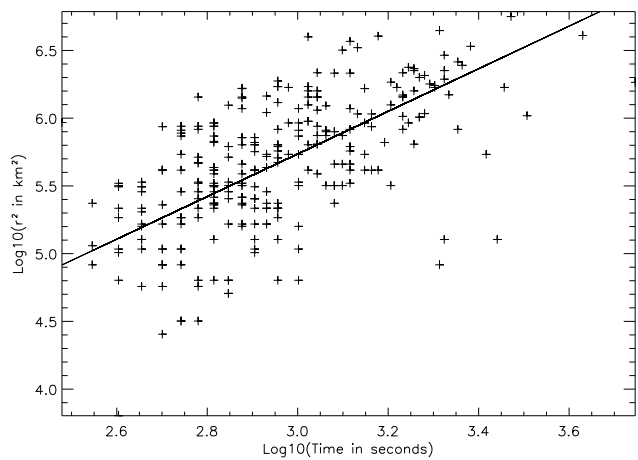}
 \caption[]{Distance of magnetic elements to their starting point
 as a function of time (log log plot).}
 \label{elgam}
\end{figure}

  The magnetic field in the quiet Sun is generally described as network (NE)  and
intranetwork (IN) components. The first represents the magnetic
field that outlines the supergranulation boundaries, the second corresponds
to small mixed-polarity elements inside the network. During the past
decade numerous works have characterized the properties
and dynamics of both IN and NE \citet{GBOKD14}. The magnetic
field emerges as ephemeral regions and weak-flux elements inside the
IN; observations here reveal migrations of small sized-elements to
supergranular boundaries.

 In this section we focus on magnetic elements detected in Stokes V images during 24 hours in the
supergranule field. The magnetic network delineates a large
supergranule that evolves continuously. The exposure time was rather
short, therefore the signal-to-noise ratio of the V signal is limited and it
was not possible to track many IN magnetic elements. An
approximate value of the magnetic field, proportional to V/I, was
derived using the usual weak-field approximation (panel 1).

Several magnetic elements (285 in total) were followed. We
studied the statistics of their lifetimes. The median of the
histogram was found at 800 s, but some elements lasted up to half an
hour. When we eliminated ephemeral elements, the data were well fit
by an exponential law of 500 s characteristic time. Figure~\ref{elvel}
displays the  velocity statistics. The median of the histogram
is 0.8 km $s^{-1}$. Velocities are more or less properly fitted by a
Maxwellian law. Figure~\ref{INelement} shows the angular distribution
of the velocity vector of the IN magnetic elements with respect to
the radial direction (from the supergranule center at~ (24~\arcsec,
39~\arcsec), showing that outward motions dominate. Most of the
velocity vectors of the IN magnetic elements ($70\%$) are measured
between  $\pm 45\degr$; this result fully corroborates those
obtained from the LCT (Fig.~\ref{ang}).

\citet{HSTS99} found evidence that in addition to the random walk
corresponding to granulation, motions of
magnetic elements are dominated by supergranulation on longer time scale. In panel 1
(magnetic field evolution) we observe both motions (random and
radial) of the IN magnetic elements, but the small number of
elements (due to the low sensitivity in Stokes V/I of our data) does
not allow us to achieve a good statistical analysis of the random
motions. Nevertheless, panel 2 clearly shows the random motions of the IN elements when 
they are located inside a TFG, and the systematic
shift (toward the borders of the supergranule) is observed when they are located 
at the edge of a TFG. This agrees with the results
obtained by \citet{OKB2012}. The IN elements are continuously
advected by large-scale convective flows, although on short
timescales the velocity is dominated by granular motions or other
effects. The highest horizontal velocities, generated by the TFG, might be 
the component that transports magnetic elements on
long timescales. These velocities exhibit a spatial
coherence at long times and large spatial scales that favor 
magnetic field diffusion on similar scales.  The motion of the most 
important TFG in the supergranule shows that five of them
are enough to build the network in six hours and then deform it the six following 
hours (see below).

The transport of the magnetic element in the photosphere is well
described by a power law $<(\Delta r)^2>=c t^\gamma$ where $<(\Delta
r)^2>$ is the mean square displacement, $c$ a constant, $t$ is the time
measured since the first detection, and $\gamma$ is the spectral
index \citet{GSBDB2014}. The distance of magnetic elements to their
starting point is plotted in Fig.~\ref{elgam} as a function of time.
The slope of the linear fit ( $\gamma$=1.57) corresponds to a
super-diffusive regime and agrees with previous results by
\citet{GSBDB2014}, \citet{GBBDS2014} and \citet{DGBC2015}. This
suggests that advection might be the dominant mechanism related to large 
spatial and long timescales  because TFG expansion produces the horizontal flow field.

\subsection{Cork evolution}

\begin{figure}
\centering
\includegraphics[width=9cm]{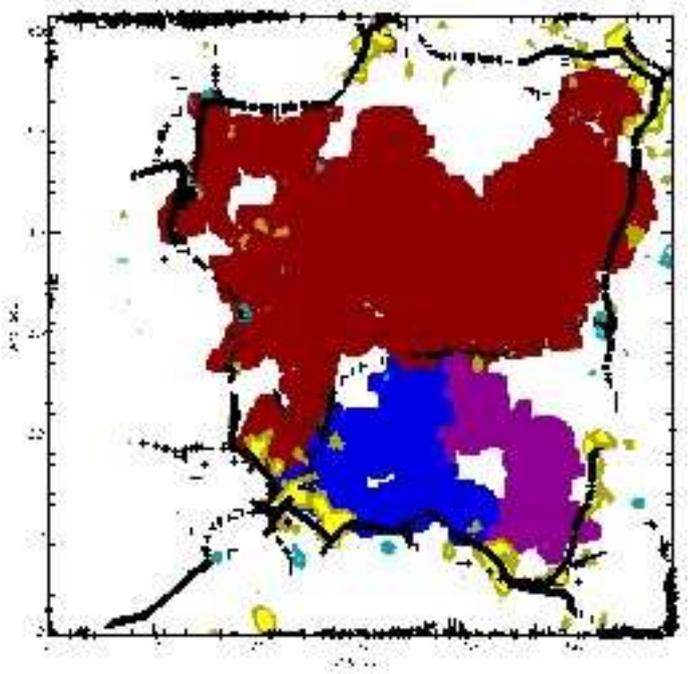}
 \caption[]{Cork locations (white cross) at the end of the 24-hour sequence in the supergranule field;
the corks are swept out at the border of the supergranule where the
magnetic field is anchored.
 The most important families (three) at the end of the sequence are overplotted 
in various colors.} \label{Corks1}
\end{figure}

\begin{figure}
\centering
\includegraphics[width=9cm]{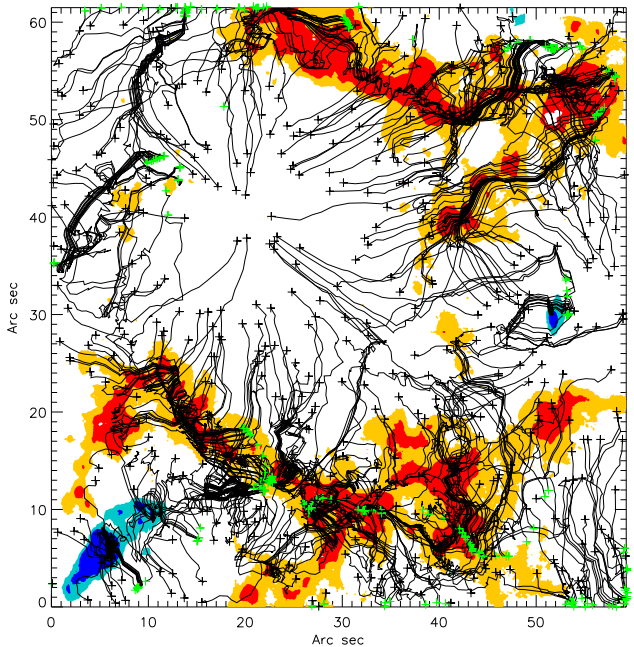}
 \caption[]{Cork trajectories during the 24-hour sequence in the supergranule field;
only a few of them are plotted for clarity.  Longitudinal
magnetic fields of the network located around the supergranule are overplotted 
in blue and red colors.} \label{Corks2}
\end{figure}

  The evolution of an initially randomly distributed passive scalar (corks)
is the best way to characterize the transport properties of the turbulent
velocity field of the solar surface. Thousands of corks  randomly distributed 
at initial time t = 0 and their surface distribution is
followed during the whole sequence as they are advected by the
horizontal velocity field. Figure~\ref{Corks1} displays the cork 
distribution that results from the TFG evolution.  This figure shows that 
a few families (three), which represent $80\%$ of the total
area, are enough to disseminate the corks to the edge of the
supergranule. The largest families contribute to the evacuation of corks
to the places where the magnetic network is located.  The combined action of two 
(or more) large families that form close in
time and space pushes corks out on tens of arcsec scales
(mesoscale). This suggests that the distribution of the magnetic
field on the solar surface depends on the most energetic TFG to form
the photospheric network or magnetic patches. Small families, which
are the most numerous, seem to play a minor role in the network
building.

Figure~\ref{Corks2} shows the cork trajectories (a selection of them for
readability) during the full sequence relative to the magnetic
network. The corks move almost radially toward the places of the
magnetic network.

 Panel 3 shows the full evolution of corks relative
to the families. We observe the rapid expulsion of corks from the
TFG and the advection of corks toward supergranule boundaries. In
addition, panel 4 shows that  during their journey corks are mainly
located inside the supergranule where the IN magnetic field is. Both
movies indicate the close link between TFG and magnetic element
evolution inside the supergranule.

\section{Link between horizontal velocities and magnetic elements}

\begin{figure}
\centering
\includegraphics[width=9cm]{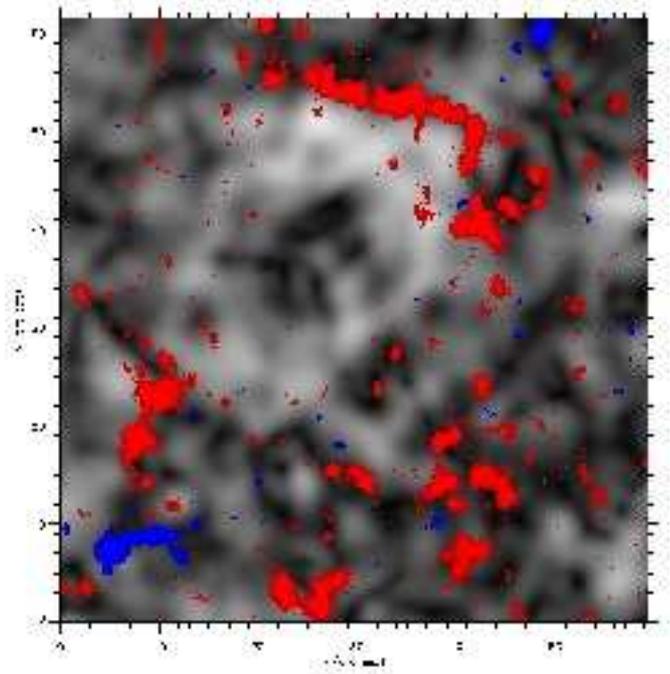}
\caption[]{{\bf Evolution of the properties of a supergranular cell. An animation is 
available online. See the appendix for details.}~ Horizontal velocity $Vh\_mag$ (gray levels) at time 422.5
min of the sequence. The longitudinal magnetic field is overplotted (blue and red).} 
\label{vit_B1}
\end{figure}

The IN flux is found to be the most important contributor to the NE
and was able to replace its entire flux in 18 to 24 hours \citet{GBOKD14}. According to 
these authors the small-scale IN elements appear as the most
permanent source of flux for the NE. The maintenance of the NE
requires the transport of the IN elements across the solar surface
toward the edges of supergranules; this process is not well documented as yet.

To obtain more details of the magnetic element transport in the quiet Sun, 
we reexamined the 2007 Hinode data \citet{RRBRM2009} with  particular 
attention to the amplitude of the horizontal velocity field. Figure~\ref{vit_B1}
(and panel 5) shows the horizontal velocity magnitude ($Vh\_mag$) and 
the longitudinal magnetic field during 24 hours, in the closed-network region (supergranule field).

The locations where the $Vh\_mag$ has the largest amplitude appear
to play an overriding role in the network formation and evolution
(deformation and localization). The movie (panel 5) reveals that the motions
of the magnetic elements inside the supergranule are driven by
$Vh\_mag$ flows regardless of their locations. In the IN, the
magnetic field elements (regardless of their polarity) are clearly
swept out (advected or diffused) toward the supergranule borders by
$Vh\_mag$ horizontal flow fronts across several arcsec with amplitudes
between 0.5 and 0.8 km/s.

\begin{figure*}
\centering
\includegraphics[width=9cm]{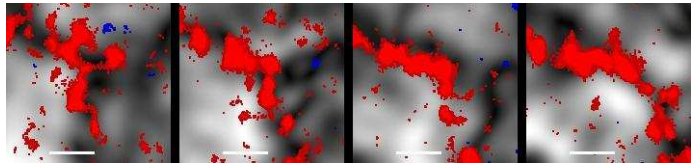}
 \caption[]{Temporal evolution of $Vh\_mag$ (gray levels) and longitudinal magnetic field (blue and red)
around coordinates (40\arcsec, 52\arcsec) (upper left corner of the supergranule field). 
From left to right times are 59 min, 169 min, 261 min and 377 min; the FOV is
$18.6~\arcsec \times 18.1~\arcsec$. The white bar gives a scale of 5~\arcsec} 
\label{V_B}
\end{figure*}

 As previously described by \citet{RRBRM2009}, the geometric
evolution of the network is linked to the TFG expansion (more
details below). Here, we observe a clear evolution of the network
shape and location also driven  by $Vh\_mag$ (Fig.~\ref{V_B}).
 We note that magnetic network patches are pushed when  there are no (or lower) 
horizontal velocities on the opposite side of the patch. If similar amplitudes of $Vh\_mag$ are
observed (right part of Fig.~\ref{V_B}, time=377 min) on both sides of
the NE, the patch remains stable (see also another example in the
bottom left part of the field in the middle of panel 5). The shape
of the network  and the flux inflow from IN to NE is strongly linked
to the $Vh\_mag$ evolution. The strongest and largest occurrences of $Vh\_mag$  
appear to be one of the most important component occurrences  to 
structure the network. The temporal coherence of $Vh\_mag$ between one to two 
hours andon a spatial scale between 3\arcsec to 12\arcsec also seems to 
play a major role. In the central part of the movie (panel 5), the
supergranule is clearly outlined by magnetic fields although in other
parts (left upper side in Fig.~\ref{vit_B1}) the network contour is
less visible because of the lower horizontal velocities. This property may be able to 
explain why the network appears more or less closed
at supergranular scale.

During the 24 hours, the supergranule delineated by the longitudinal
magnetic field appears to move toward the right side of the FOV. 
This is quite surprising because a detailed check of our data
alignment did not show any drift, whereas the NE appears to rotate
6\% faster. We observe, as \citet{LGB2015}, a greater presence of
the NE in the western part (Fig.~\ref{vit_B1}) of the supergranule.

\section{Discussion}

The origin and detailed evolution of the supergranulation is still 
an open question. The supergranular flow scale is unexplained.
The length scale (30 Mm) and lifetime (24 to 48 hours) promote
studies with low resolutions \citet{DET04}. Statistical averaging
over measurements of numerous supergranules (\citealt{DJ2015},
\citealt{DHC2014}, \citealt{LGB2015}) allows us to describe global 
characteristics, but it is difficult to identify the mechanism of
supergranule formation because a correspondence between average
quantities and the high temporal and spatial processes is missing.

 Supergranules are often described as convective structures
 with an upflowing central region, horizontally diverging flows,
 and downflows at the edge. Because downflows are
co-spatial with magnetic elements, many authors have argued
that supergranulation is a magnetoconvection phenomenon
\citet{BEFP1999}. However, simulations do not agree on the
causal mechanism and helium ionization is invoked \citet{HS2014}.

Supergranulation is often described as an isotropic flow diffusing
magnetic fields toward their periphery. From an observational point
of view, the large-scale structure is inhomogeneous in size,
velocity magnitude and shape (see Fig. 3 of \citealt{RR10} issued
from the largest high-resolution FOV of horizontal velocities up to
now). This reveals the anisotropic character of flows, which has
consequences on the diffusion of magnetic fields across the solar
surface on a long timescale. The NE outlines supergranular cells
\citet{RUT1999} but is often incomplete at the boundaries. The
magnetic NE is quite fragmented and exhibits many patches of strong
fields coinciding with vertices where supergranular flows meet,
presumably marking the sites of downflows required by mass
conservation \citet{HSTS99}. This also raises the question of the
link between flows and NE formation. In addition, recent works using
helioseismic data emphasized the difficulty of deriving the formation 
height of the supergranulation, as well as the velocity profile with
altitude \citet{DJ2015}. This means that the supergranular model with warmer
fluid upflowing at the center and cooler fluid downflowing at the
boundaries is not supported by observations (see also
\cite{RAST03}).

The high spatial and temporal resolution of Hinode/SOT data allowed us to
study great detail the evolution of the photospheric
flows relative to TFG that cover the solar surface for 24 hours. Large TFGs,
with sizes close to supergranulation, are not numerous but are
sufficient to structure the flows. TFGs grow in area by a succession
of granule explosions that are correlated in time. TFGs and their
mutual interactions are able to sweep out magnetic elements to the
border of supergranules. Almost simultaneous explosions of granules
at TFG birth generate coherent flow fronts with large spatial
scales (3 to 12\arcsec) and long lifetimes (one to two hours) with
magnitudes in the range 0.5 to 0.8 km/s. New branches of TFGs during
their life generate collective effects through horizontal velocities 
that supply the general flow. Flow fronts propagate when there is
no counterpart motion resulting from interactions with other TFG.
Horizontal flows tend to increase close to the limits of expanding
TFGs, extracting out magnetic elements more or less radially toward the
borders of supergranules. The occurrence of large-scale velocity
maxima is probably one of the main phenomena that diffuse (or
advect) the magnetic field and also affect the NE shape and
evolution. Magnetic patches in the IN are found at the edge of
velocity fronts, indicating a possible link between IN and TFG
evolution. The lack of the magnetic signal observed at the center
of supergranules \citet{Stang14} might be related to the formation
and evolution of TFGs that quickly expels corks that are considered to be 
proxies of magnetic flux tubes.

 The velocity fronts produced by the TFG may not only contribute to
magnetic diffusion in supergranules but also be generated by small-scale
dynamo \citet{TC2008}. Tobias and Cattaneo showed that the magnetic field 
amplification depends on spatially and temporally coherent structures.

 We observed an intriguing event in the evolution of the NE : while data alignment
was perfectly controlled during 24 hours (solar rotation and
satellite shifts where corrected for), the NE moved to the right side of the
FOV (6\% faster westward). This is quite difficult to interpret in
the context of the classical convective approach of the
supergranulation. Motions of NE patches appear to be governed by the
largest TFG expansion and the resulting horizontal flows. We suggest
that NE location and shape are controlled by TFG evolution in space
and time. The new question arising from our analysis is whether TFGs
are only the basic elements of supergranules or the main
structure at the origin of NE formation and evolution. We cannot
answer this question; numerical simulations could help to
distinguish between these hypotheses, however.

\section{Conclusions}

 Understanding the mechanism that diffuses magnetic fields is still a challenge.
 This important, because the NE contribution to the global solar magnetism is
 known to be similar to the flux of active regions at maximum
 activity.

 From our analysis at high spatial and temporal resolution during 24 hours, TFGs appear as
one of the main elements of supergranules which diffuse and advect
magnetic fields over the solar surface. We discovered that the largest
TFG evolutions and mutual interactions produce cumulative effects that are able 
to build horizontally coherent flows with longer lifetimes than
granulation (one to two hours) and with scale lengths of up to 12\arcsec. We
observed that TFGs grow in area by a succession of granule
explosions, which are correlated in time, expelling corks on the 
typical size of the mesoscale. Diverging flows at the beginning of
TFG life generate horizontal and outward velocity fronts with
magnitudes between 0.5 and 0.8 km/s over several arcsec.

These coherent flows are often located at TFG edges and can propagate
in the direction where they do not need to compete with another large
TFG. The flows act on the location and shape of the NE and are
compatible with the recent work of \citealt{GBOKD14} who indicated
that the IN might supply the magnetic flux in the NE in only
9 to 13 hours. We suggest that supergranular NE might be a spatial
pattern that originates from the action of TFGs on the scale of
supergranules.

The rotation of the NE (6\% faster westward) found in our data
raises the question of the role of TFGs on the solar surface. TFG
flows might be the origin of NE formation by feeding and shaping it.
We must now test this conjecture by analyzing simulations with large
FOVs with known parameters to determine whether TFGs  are able to provide 
sufficient diffusion rates to account for large-scale distribution of the
magnetic fields in the solar photosphere \citet{STW95}.

\begin{acknowledgements}
 We thank the \textit{Hinode}/SOT team for assistance in acquiring and
processing the data.  \textit{Hinode} is a Japanese mission developed
and launched by ISAS/JAXA, collaborating with NAOJ as a domestic
partner, NASA and STFC (UK) as international partners. Support for
the post-launch operation is provided by JAXA and NAOJ (Japan), STFC
(U.K.), NASA, ESA, and NSC (Norway). This work was granted access to the
HPC resources of CALMIP under the allocation 2011-[P1115].  The authors 
wish to thank the anonymous referee for very helpful comments and 
suggestions that improved the quality of the manuscript.
\end{acknowledgements}

\bibliographystyle{aa}
\bibliography{biblio}

\begin{thebibliography}{28}
\expandafter\ifx\csname natexlab\endcsname\relax\def\natexlab#1{#1}\fi

\bibitem[{{Berrilli} {et~al.}(1999){Berrilli}, {Ermolli}, {Florio}, \&
  {Pietropaolo}}]{BEFP1999}
{Berrilli}, F., {Ermolli}, I., {Florio}, A., \& {Pietropaolo}, E. 1999, \aap,
  344, 965

\bibitem[{{Crouch} {et~al.}(2007){Crouch}, {Charbonneau}, \&
  {Thibault}}]{CCT07}
{Crouch}, A.~D., {Charbonneau}, P., \& {Thibault}, K. 2007, \apj, 662, 715

\bibitem[{{DeGrave} \& {Jackiewicz}(2015)}]{DJ2015}
{DeGrave}, K. \& {Jackiewicz}, J. 2015, \solphys, 290, 1547

\bibitem[{{Del Moro} {et~al.}(2015){Del Moro}, {Giannattasio}, {Berrilli},
  {Consolini}, {Lepreti}, \& {Go{\v s}i{\'c}}}]{DGBC2015}
{Del Moro}, D., {Giannattasio}, F., {Berrilli}, F., {et~al.} 2015, \aap, 576,
  A47

\bibitem[{{DeRosa} \& {Toomre}(2004)}]{DET04}
{DeRosa}, M.~L. \& {Toomre}, J. 2004, \apj, 616, 1242

\bibitem[{{Duvall} {et~al.}(2014){Duvall}, {Hanasoge}, \&
  {Chakraborty}}]{DHC2014}
{Duvall}, T.~L., {Hanasoge}, S.~M., \& {Chakraborty}, S. 2014, \solphys, 289,
  3421

\bibitem[{{Giannattasio} {et~al.}(2014{\natexlab{a}}){Giannattasio},
  {Berrilli}, {Biferale}, {Del Moro}, {Sbragaglia}, {Bellot Rubio}, {Go{\v
  s}i{\'c}}, \& {Orozco Su{\'a}rez}}]{GBBDS2014}
{Giannattasio}, F., {Berrilli}, F., {Biferale}, L., {et~al.}
  2014{\natexlab{a}}, \aap, 569, A121

\bibitem[{{Giannattasio} {et~al.}(2014{\natexlab{b}}){Giannattasio},
  {Stangalini}, {Berrilli}, {Del Moro}, \& {Bellot Rubio}}]{GSBDB2014}
{Giannattasio}, F., {Stangalini}, M., {Berrilli}, F., {Del Moro}, D., \&
  {Bellot Rubio}, L. 2014{\natexlab{b}}, \apj, 788, 137

\bibitem[{{Go{\v s}i{\'c}} {et~al.}(2014){Go{\v s}i{\'c}}, {Bellot Rubio},
  {Orozco Su{\'a}rez}, {Katsukawa}, \& {del Toro Iniesta}}]{GBOKD14}
{Go{\v s}i{\'c}}, M., {Bellot Rubio}, L.~R., {Orozco Su{\'a}rez}, D.,
  {Katsukawa}, Y., \& {del Toro Iniesta}, J.~C. 2014, \apj, 797, 49

\bibitem[{{Hagenaar} {et~al.}(1999){Hagenaar}, {Schrijver}, {Title}, \&
  {Shine}}]{HSTS99}
{Hagenaar}, H.~J., {Schrijver}, C.~J., {Title}, A.~M., \& {Shine}, R.~A. 1999,
  \apj, 511, 932

\bibitem[{{Hanasoge} \& {Sreenivasan}(2014)}]{HS2014}
{Hanasoge}, S.~M. \& {Sreenivasan}, K.~R. 2014, \solphys, 289, 3403

\bibitem[{{Hart}(1954)}]{H54}
{Hart}, A.~B. 1954, \mnras, 114, 17

\bibitem[{{Langfellner} {et~al.}(2015){Langfellner}, {Gizon}, \&
  {Birch}}]{LGB2015}
{Langfellner}, J., {Gizon}, L., \& {Birch}, A.~C. 2015, \aap, 579, L7

\bibitem[{{Leighton} {et~al.}(1962){Leighton}, {Noyes}, \& {Simon}}]{LNS62}
{Leighton}, R.~B., {Noyes}, R.~W., \& {Simon}, G.~W. 1962, \apj, 135, 474

\bibitem[{{November}(1989)}]{Nov1989}
{November}, L.~J. 1989, \apj, 344, 494

\bibitem[{{Orozco Su{\'a}rez} {et~al.}(2012{\natexlab{a}}){Orozco Su{\'a}rez},
  {Bellot Rubio}, \& {Katsukawa}}]{OBK2012}
{Orozco Su{\'a}rez}, D., {Bellot Rubio}, L.~R., \& {Katsukawa}, Y.
  2012{\natexlab{a}}, in Astronomical Society of the Pacific Conference Series,
  Vol. 463, Second ATST-EAST Meeting: Magnetic Fields from the Photosphere to
  the Corona., ed. T.~R. {Rimmele}, A.~{Tritschler}, F.~{W{\"o}ger},
  M.~{Collados Vera}, H.~{Socas-Navarro}, R.~{Schlichenmaier}, M.~{Carlsson},
  T.~{Berger}, A.~{Cadavid}, P.~R. {Gilbert}, P.~R. {Goode}, \&
  M.~{Kn{\"o}lker}, 57

\bibitem[{{Orozco Su{\'a}rez} {et~al.}(2012{\natexlab{b}}){Orozco Su{\'a}rez},
  {Katsukawa}, \& {Bellot Rubio}}]{OKB2012}
{Orozco Su{\'a}rez}, D., {Katsukawa}, Y., \& {Bellot Rubio}, L.~R.
  2012{\natexlab{b}}, \apjl, 758, L38

\bibitem[{{Rast}(2003)}]{RAST03}
{Rast}, M.~P. 2003, \apj, 597, 1200

\bibitem[{{Rieutord} \& {Rincon}(2010)}]{RR10}
{Rieutord}, M. \& {Rincon}, F. 2010, Living Reviews in Solar Physics, 7, 2

\bibitem[{{Roudier} {et~al.}(2003){Roudier}, {Ligni{\`e}res}, {Rieutord},
  {Brandt}, \& {Malherbe}}]{RLRBM03}
{Roudier}, T., {Ligni{\`e}res}, F., {Rieutord}, M., {Brandt}, P.~N., \&
  {Malherbe}, J.~M. 2003, \aap, 409, 299

\bibitem[{{Roudier} {et~al.}(2009){Roudier}, {Rieutord}, {Brito}, {Rincon},
  {Malherbe}, {Meunier}, {Berger}, \& {Frank}}]{RRBRM2009}
{Roudier}, T., {Rieutord}, M., {Brito}, D., {et~al.} 2009, \aap, 495, 945

\bibitem[{Roudier {et~al.}(1999)Roudier, Rieutord, Malherbe, \&
  Vigneau}]{RRMV99}
Roudier, T., Rieutord, M., Malherbe, J., \& Vigneau, J. 1999, \aap, 349, 301

\bibitem[{{Rutten}(1999)}]{RUT1999}
{Rutten}, R.~J. 1999, in Astronomical Society of the Pacific Conference Series,
  Vol. 184, Third Advances in Solar Physics Euroconference: Magnetic Fields and
  Oscillations, ed. B.~{Schmieder}, A.~{Hofmann}, \& J.~{Staude}, 181--200

\bibitem[{{Simon} {et~al.}(1995){Simon}, {Title}, \& {Weiss}}]{STW95}
{Simon}, G.~W., {Title}, A.~M., \& {Weiss}, N.~O. 1995, \apj, 442, 886

\bibitem[{{Stangalini}(2014)}]{Stang14}
{Stangalini}, M. 2014, \aap, 561, L6

\bibitem[{{Stein} {et~al.}(2009){Stein}, {Nordlund}, {Georgoviani}, {Benson},
  \& {Schaffenberger}}]{Stein2009}
{Stein}, R.~F., {Nordlund}, {\AA}., {Georgoviani}, D., {Benson}, D., \&
  {Schaffenberger}, W. 2009, in Astronomical Society of the Pacific Conference
  Series, Vol. 416, Solar-Stellar Dynamos as Revealed by Helio- and
  Asteroseismology: GONG 2008/SOHO 21, ed. M.~{Dikpati}, T.~{Arentoft},
  I.~{Gonz{\'a}lez Hern{\'a}ndez}, C.~{Lindsey}, \& F.~{Hill}, 421

\bibitem[{{Suematsu} {et~al.}(2008){Suematsu}, {Tsuneta}, {Ichimoto},
  {Shimizu}, {Otsubo}, {Katsukawa}, {Nakagiri}, {Noguchi}, {Tamura}, {Kato},
  {Hara}, {Kubo}, {Mikami}, {Saito}, {Matsushita}, {Kawaguchi}, {Nakaoji},
  {Nagae}, {Shimada}, {Takeyama}, \& {Yamamuro}}]{STISO08}
{Suematsu}, Y., {Tsuneta}, S., {Ichimoto}, K., {et~al.} 2008, \solphys, 249,
  197

\bibitem[{{Tobias} \& {Cattaneo}(2008)}]{TC2008}
{Tobias}, S.~M. \& {Cattaneo}, F. 2008, Physical Review Letters, 101, 125003

\end{thebibliography}

\Online

\begin{appendix}

\begin{figure*}
\centering
\includegraphics[width=18cm]{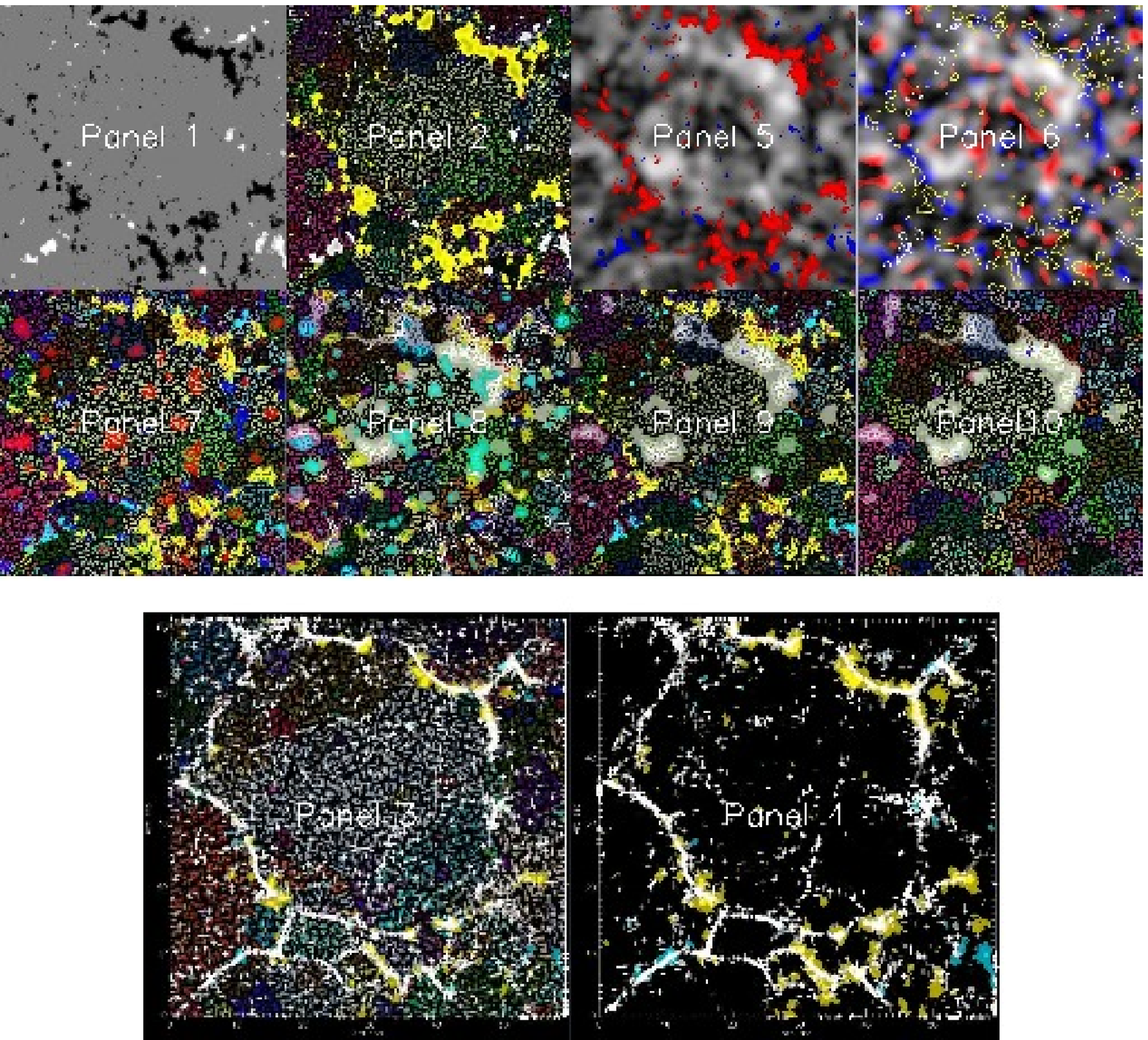}
\caption[]{ }
\label{appendix}
\end{figure*}

{\bf Evolution of a supergranule}.

{\bf Online supplement material:  movie12-510.mp4 and movie34.mp4 (Mpeg 4 format)}.

 In this online material we show the evolution of the magnetic field and several derived properties 
as discussed in the main text. A snapshot of the movie is show in Fig A.1. The field of view 
(60~\arcsec x 62~\arcsec) and is centered on the well-formed superganule with an almost 
closed magnetic network at the boundaries. The time step is 50 s between each frame and the duration 
is 24 hours (1716 frames). The pixel size is 0.16~\arcsec. The contents of the individual panels 
is as follows:

{\bf Panel 1:} Evolution of the magnetic field .

{\bf Panel 2:}  Evolution of the TFG with various colors; 
          the longitudinal magnetic field is superimposed (yellow and white).

{\bf Panel 3:} Evolution of corks (indicated by crosses) relative
to the TFG (various colors); the longitudinal magnetic field is
superimposed (yellow and blue). 

{\bf Panel 4:} Evolution of corks (indicated by crosses) relative
to the intranework (IN) and network (NE) longitudinal magnetic field
(yellow and blue).

{\bf Panel 5:} Evolution of horizontal velocity magnitudes
($Vh\_mag$, gray levels) and the longitudinal magnetic field (red
and blue).

{\bf Panel 6:} Evolution of horizontal velocity magnitudes
$Vh\_mag$, divergence of horizontal velocities (positive for
divergent in red, negative for convergent in blue) together with
magnetic fields (yellow and white contours).

{\bf Panel 7:}  Evolution of the positive divergence (in red)
relative to the families (TFG in various colors) and magnetic
fields (yellow and blue).

{\bf Panel 8:} Evolution of the positive divergence (in blue)
relative to the families (TFG in various colors), magnetic field
magnitude (white contours) and horizontal velocity magnitudes
$Vh\_mag$ (gray levels).

{\bf Panel 9:} Evolution of the families (TFG in various colors),
horizontal velocity magnitudes $Vh\_mag$ (white) and magnetic fields
(yellow and blue contours).

{\bf Panel 10:}  Evolution of the families (TFG in various colors)
and horizontal velocity magnitudes $Vh\_mag$ (white).

{\bf The movies can be found in full resolution at the address:
http://www.lesia.obspm.fr/perso/jean-marie-malherbe/Hinode2007/hinode2007.html}

\end{appendix}

\end{document}